\documentclass[aip,jmp,10pt]{revtex4-1}
\usepackage{amssymb,amsmath}

\usepackage{mathrsfs}
\usepackage{epsfig}
\usepackage{epstopdf}

\begin{document}

\title{Noether's Theorem for Dissipative Quantum Dynamical Semigroups}
\author{John E. Gough} \email{jug@aber.ac.uk}
\affiliation{Aberystwyth University, Aberystwyth, SY23 3BZ, Wales, United Kingdom}

\author{Tudor S. Ratiu} \email{tudor.ratiu@epfl.ch}
\affiliation{Section de Math\'{e}matiques and Bernoulli Center, \'{E}cole Polytechnique F\'{e}d\'{e}rale de Lausanne,\\ Lausanne, CH 1015 Switzerland.}

\author{Oleg. G. Smolyanov} \email{smolyanov@yandex.ru}
\affiliation{Mechanics and Mathematics Faculty, Moscow State University, Moscow, 119991 Russia.}

\begin{abstract}
Noether's Theorem on constants of the motion of dynamical systems has recently been extended to classical dissipative systems (Markovian semi-groups) by Baez and Fong \cite{Baez_Fong}. We show how to extend these results to the fully quantum setting of quantum Markov dynamics. For finite-dimensional Hilbert spaces, we construct a mapping
from observables to CP maps that leads to the natural analogue of their criterion of commutativity with the infinitesimal generator of the Markov dynamics. Using
standard results on the relaxation of states to equilibrium under quantum dynamical semi-groups, we are able to characterise the constants of the motion under 
quantum Markov evolutions in the infinite-dimensional setting under the usual assumption of existence of a stationary strictly positive density matrix.
In particular, the Noether constants are identified with the fixed point of the Heisenberg picture semigroup.
\end{abstract}

\maketitle

\section{Introduction}
Noether's Theorem has traditionally been formulated within the framework of closed systems where it has been of central importance in formulating
the concept of constants of the motion for Lagrangian dynamical systems, however, the widespread belief is that it does not apply to dissipative systems.
Recently, a version of the Theorem has been formulated by Baez and Fong \cite{Baez_Fong} for classical Markovian models. Here a constant of the motion 
is a random variable on the state space of a classical Markov process
whose probability distribution is time invariant under the Markov semigroup. While entirely classical, their investigation draws on analogies to
closed quantum dynamics: indeed random variables may be identified with the operators 
of pointwise multiplication by the random functions, and they establish that constancy of the motion commutativity of these operators with the 
Markov transition mechanism operators (or their infinitesimal generator).

In this paper, we extend these results to the setting quantum dynamical semigroups \cite{Alicki_Fannes}. 

\section{Classical Markov Processes with Finite State Spaces}
This section reviews the results of Baez and Fong \cite{Baez_Fong}. Our only contribution at this stage is to present the results in a fashion that makes it easier to see extensions to quantum Markov models in the following sections.

Let $\Gamma $ be a finite set, say, $\Gamma =\left\{ 1,\cdots ,d\right\} $ for definiteness. 
We denote by $\Sigma \left( \Gamma \right) $ the set of all probability vectors $\mathbf{p}$, that is $d$-tuples  
$ \mathbf{p}=\left[ p_x \right]$,  with $p_{x}\geq 0$, and $\sum_{x}p_{x}=1$.

A \textit{classical dynamical semigroup} (CDS) with state space $\Gamma$ is determined by a transition mechanism $\left( T_{t}\right) _{t\geq 0}$ of the form 
$T_{t}=\left[ T_{xy} (t) \right] $,
that forms a matrix semigroup: $T_{t}T_{s}=T_{t+s}$, with $T_{0}=I_{d}$, the $d\times d$ identity matrix, such that $T\left( t\right) $ maps $\Sigma \left( \Gamma \right) $ to itself. The entries $T_{xy}(T)$ give the conditional probabilities for a transition from state $y$ to state $x$ in time $t$. 
An initial probability vector $\mathbf{p}$ evolves under the transition mechanism as 
$\mathbf{p}\left( t\right) =T_{t}\,\mathbf{p} $.
Note that we require the identities 
\begin{eqnarray}
\sum_{x}T_{xy}\left( t\right) =1,\quad \text{ for all }y\in \Gamma .
\label{eq:T-identity}
\end{eqnarray}

It follows from the semigroup law that 
\begin{eqnarray}
T_{t}=e^{tM}
\end{eqnarray}
where $M\in \mathbb{R}^{d\times d}$ is called the infinitesimal generator of the transition mechanism. To lowest order in $t$ we have  
$0\leq p_{x}\left( t\right) =p_{x}+t\sum_{y}M_{xy}p_{k}+O\left( t^{2}\right)$
and to be true for all $t\geq 0$ we require that 
\begin{eqnarray}
M_{xy}\geq 0,\quad \text{for all } x\neq y.  \label{eq:M positive of diagonal}
\end{eqnarray}
We also require that 
\begin{eqnarray}
\sum_{x\in \Gamma }M_{xy}=0,\quad \text{\ for all }y\in \Gamma ,  \label{eq:M identity}
\end{eqnarray}
which is the infinitesimal form of (\ref{eq:T-identity}). The requirements (\ref{eq:M positive of diagonal}) 
and (\ref{eq:M identity}) characterise the infinitesimal generators of Markov transition matrices.

Now let $A$ be a random variable on $\Gamma $, its expectation for a fixed probability vector $\mathbf{p}\in \Sigma \left( \Gamma \right) $ is 
\begin{eqnarray*}
\mathbb{E}\left[ A\right] =\sum_{x\in \Gamma }A\left( x\right) \,p_{x} .
\end{eqnarray*}
This may be written as $\mathbb{E}\left[ A\right] =\sum_{a}a\,\mathbb{K}^{A}(a)$ where
\begin{eqnarray}
\mathbb{K}^{A}\left( a\right) =\sum_{\left\{ x\in \Gamma:A(x)=a\right\} }p_{x}=\Pr \left\{ A=a\right\} . 
\end{eqnarray}
$\mathbb{K}^{A}$ is the probability distribution of $A$ determined by $\mathbf{p}$. The probability
distribution determined by $\mathbf{p}\left( t\right) =T\left( t\right) \mathbf{p}$ is similarly denote as $\mathbb{K}_{t}^{A}$
\begin{eqnarray*}
\mathbb{E}_{t}\left[ f(A)\right] =\sum_{x,y\in \Gamma }f(A\left( x\right)
)\,T_{xy}\left( t\right) \,p_{y}\equiv \sum_{a}f\left( a\right) \,\mathbb{K}_{t}^{A}\left( a\right) .
\end{eqnarray*}

\bigskip

\noindent \textbf{Definition}
\textit{
A random variable $A$ with sample space $\Gamma $ is said to be a constant for the CDS with Markov mechanism $T$ on $\Gamma$ if, 
for each initial $\mathbf{p}\in \Sigma \left( \Gamma \right) $, its probability distribution $\mathbb{K}_{t}^{A}$
does not depend on $t\geq 0$.}

\bigskip

This means that, for each $\mathbf{p}\in \Sigma \left( \Gamma \right) $ and any polynomial function $f$, we have for all $t\geq 0$
\begin{eqnarray*}
\mathbb{E}_{t}\left[ f(A)\right] =\mathbb{E}\left[ f(A)\right] .
\end{eqnarray*}

\bigskip

The transition mechanism furnishes an equivalence relation on $\Gamma $ as follows: we say that $x\sim y$ if and only if we have the matrix element $\left[ M^{n}\right] _{xy}\neq 0$ for some positive integer power $n$. We
shall denote the $\sigma $-algebra \cite{sigma-algebra}
 generated by the equivalence classes determined by transition mechanism by $\mathscr{M}$.

\bigskip

\noindent\textbf{Theorem  (Baez-Fong) \cite{Baez_Fong}} 
\textit{
For each random variable $A$ on $\Gamma $, we define the $d\times d$ matrix 
\begin{eqnarray}
\hat{A} &=&diag\left( A\left( 1\right) ,\cdots ,A\left( d\right) \right)  =\left[ 
\begin{array}{cccc}
A\left( 1\right)  & 0 & \cdots  & 0 \\ 
0 & A(2) &  & 0 \\ 
\vdots  &  & \ddots  & \vdots  \\ 
0 & 0 & \cdots  & A(d)
\end{array}
\right] .
\label{eq:BFhat}
\end{eqnarray}
Let $M$ be the infinitesimal generator for a Markov transition mechanism on $\Gamma $. Then the following are equivalent:}
\begin{enumerate}
\item  \textit{$A$ is a constant of the CDS;}

\item  \textit{The mean and variance of
the probability distribution $\mathbb{K}_{t}^{A}$ are constant in time;}

\item  \textit{$A$ is measurable with respect to the $\sigma $-algebra $\mathscr{M}$;}

\item  $\left[ \hat{A},M\right] =0;$
\end{enumerate}

The proof can be found in \cite{Baez_Fong}, and we sketch the arguments for completeness.

\textit{(1) implies (2)} Clearly constancy of the distribution implies
constancy of the first and second moments.

\textit{(2) implies (3)} If (2) holds then the time-derivative of the expectation  $ \sum_{x,y} A(x)^m T_{xy} (t) p(y) $ vanishes for every probability vector and 
for $m=0,1,2$. This implies that $ \sum_{x \in \Gamma} A(x)^m M_{xy}=0$ for all $y\in \Gamma$ and for $m=0,1,2$. In turn,  (by expanding the bracket) the expression
$\sum_{xy} [A(x)-A(y) ]^2 \, M_{xy}$ vanishes identically. As $M_{xy} \geq$ for$x \neq y$, we see that if $M_{xy} \neq 0$  and $x\neq y$, then $A(x)-A(y)=0$. By
transitivity of the equivalence relation, we get that $A(x) = A(y)$ for all $x \sim y$.

\textit{(3) implies (4)} We note that the matrix $C=[ \hat{A}, M]$
has components $C_{xy} = A(x)\, M_{xy } - M_{xy} \, A(y)$. Evidently we have $C_{xy}=0$ whenever $x$ and $y$ belong to different equivalence classes, since we then have 
$M_{xy} =0$. If we additionally assume (3) that $A(x)=A(y)$ whenever $x\sim y$, we see immediately that $C_{xy} =[A(x)-A(y) ] \, M_{xy} \equiv 0$ whenever $x\sim y$.

\textit{(4) implies (1)} We see that $ \frac{d}{dt} \sum_x f(A(x)) \, p_x (t) =\sum_{x,y} f(A(x) \, M_{xy} \, p_y (t)$, however if (4) holds, then we have the identity
$f(A(x))\, M_{xy } = M_{xy} \, f( A(y))$, and so
$ \frac{d}{dt} \sum_x f(A(x)) \, p_x (t) =\sum_{x,y} f(A(y) \, M_{xy} \, p_y (t)$, but this will vanish identically due to (\ref{eq:M identity}).

\bigskip

More generally, let $( \Gamma , \mathcal{G} , \mu )$ be a $\sigma$-finite measure space and set
\[
\mathfrak{A} = L^\infty ( \Gamma , \mathcal{G} , \mu ).
\]
The space of densities $L^1 ( \Gamma , \mathcal{G} , \mu )$ is the dual to $\mathfrak{A}$ and a semigroup $(T_t)_{t\geq 0}$ on the densities leads to a dual 
semigroup $(J_t)_{t \geq 0}$ on $\mathfrak{A}$ given by
\[
\int_\Gamma A(x) \, (T_t S) (x) \, \mu[dx ]= \int_\Gamma  ( J_t A)(x) \, S( x) \, \mu [dx] ,
\]
for all $A\in \mathfrak{A}, \, S \in L^1 ( \Gamma , \mathcal{G} , \mu )$. With an obvious abuse of notation, we write $ (T_t \mathbb{P}) [dx]$ for $ (T_t \rho) (x) \,
\mu [dx]$ whenever $\mathbb{P}$ is absolutely continuous with respect to $\mu$ with Randon-Nikodym derivative $\rho$. 
The probability distribution of a random variable $A \in \mathfrak{A}$
at time $t$ for a given initial distribution $\mathbb{P} [dx] = \rho (x) \, \mu [dx]$ is then
\[
\mathbb{K}^{A, \rho}_t = (T_t  \mathbb{P}) \circ A^{-1 },
\]
and the random variable $A$ is a constant under the CDS if $\mathbb{K}^{A, \rho}_t$ is independent of $t$ for all fixed probability densities 
$\rho \in L^1 ( \Gamma , \mathcal{G} , \mu )$.

Baez and Fong establish the more general result for a continuous CDS $(T_t)_{t\geq 0}$ on $L^1 ( \Gamma , \mathcal{G} , \mu )$,
that a random variable $A$ satisfies $[ \hat {A} , T_t ]=0$ (for all $t \geq 0$) if
and only if the first two moments of $\mathbb{K}^{A, \rho}_t$ are independent of $t$ for all fixed probability densities 
$\rho \in L^1 ( \Gamma , \mathcal{G} , \mu )$.

It is instructive to express these results in a more algebraic language so as to anticpate the quantum version. The collection of random variables
$\mathfrak{A} = L^\infty ( \Gamma , \mathcal{G} , \mu )$ is in fact a commutative von Neumann algebra. 
The random variables that are the constants under a given Markovian dynamics are, in the case of finite sample space $\Gamma$ at any rate,
identified as the $\mathscr{M}$-measurable ones. Mathematically there is a one-to-one 
correspondence between $\sigma$-algebras $\mathscr{M}$ and the algebras of  bounded $\mathscr{M}$-measurable functions $\mathfrak{M}$. In fact,
$\mathfrak{M}$ will then be a commutative von Neumann sub-algebra, so an alternative statement of the Baez-Fong result is that a bounded random variable will be a constant of a classical Markov semigroup if and only if it belongs to the von Neumann algebra $\mathfrak{M}$ determined by the transition mechanism (this applies to their result on general Markov 
semigroups).

\bigskip

\section{Noether's Theorem for Quantum Markov Dynamics (Finite State Space)}

We now move to the setting of a finite-dimensional Hilbert space $\mathfrak{h}=\mathbb{C}^{d}$. The space of states becomes the set $\Sigma _{d}$ of density matrices, that is, matrices $\varrho \in \mathbb{C}^{d\times d}$
satisfying the properties $\varrho \geq 0$ and $\mathrm{tr}\left\{ \varrho \right\} =1 $. Each $\varrho \in \Sigma _{d}$ determines the expectation (or state)
\begin{eqnarray*}
\mathbb{E}\left[ \, \cdot \,  \right] =\mathrm{tr}\left\{ \varrho \,\cdot \right\} .
\end{eqnarray*}

Observable quantities are described by observables, that is Hermitean $d\times d$ matrices. Each observable $A$ admits a spectral decomposition $A=\sum_{a}a\,P_{a}$ where the sum is over the spectrum of $A$ and $P_{a}$ is
the orthogonal projection onto the eigenspace of $a$ corresponding to eigenvalue $a$. We find that $\mathbb{E}\left[ f(A)\right] =\sum_{a}f(a)\,\mathbb{K}^{A}(a)$ where the distribution $\mathbb{K}^{A}$ is now given by
Born's rule $\mathbb{K}^{A}(a)=\mathrm{tr}\left\{ \varrho P_{a}\right\} $. 

Let $\mathfrak{A}=\mathfrak{B}\left( \mathfrak{h}\right) $ be the algebra of bounded operators on the Hilbert space, here just represented as $d\times d$ complex valued matrices. We are interested in linear maps $\mathcal{T}:\mathfrak{A}\mapsto \mathfrak{A} $ which take density matrices to density matrices. Furthermore, we impose the standard requirement that the maps are \textit{completely positive} (CP), see e.g. \cite{Alicki_Fannes}. They should also be trace-preserving:
\begin{eqnarray*}
\mathrm{tr}\left\{ \mathcal{T}\left( S\right) \right\} =\mathrm{tr}\left\{ S\right\} ,\quad \text{for
all }S\in \mathfrak{A}\text{.}
\end{eqnarray*}
As is well-known, such CP maps (also called quantum communication channels) may be represented as \cite{Alicki_Fannes}
\begin{eqnarray*}
\mathcal{T}\left( S\right) =\sum_{k}V_{k}SV_{k}^{\ast }
\end{eqnarray*}
where $\left\{ V_{k}\right\} _{k}$ is a set of operators in $\mathfrak{A}$
called the Kraus maps. The trace preserving property is ensured if $%
\sum_{k}V_{k}^{\ast }V_{k}=I_{d}$.

A family $\left( \mathcal{T}_{t}\right) _{t\geq 0}$ of CP trace-preserving maps
forming a semigroup $\left( \mathcal{T}_{0}=id,\mathcal{T}_{t}\circ \mathcal{T}_{s}=\mathcal{T}_{t+s}\right) $
is said to be a \textit{quantum dynamical semigroup} (QDS).

\bigskip 

\noindent \textbf{Definition}
\textit{An observable $A$ is a constant of QDS if, for any
initial state $\varrho \in \Sigma _{d}$, we have $\mathbb{K}%
_{t}^{A}(a)=\mathrm{tr}\left\{ \mathcal{T}_{t}(\varrho )\,P_{a}\right\} $ independent of $t\geq
0$.}

\bigskip 

Let $A\in \mathfrak{A}$ be an observable with spectral decomposition $%
\sum_{a}a\,P_{a}$. For any polynomial $f$, we define a map $\widehat{f\left(
A\right) }:\mathfrak{A}\mapsto \mathfrak{A}$ by
\begin{eqnarray}
\widehat{f\left( A\right) }\left( S\right) \triangleq \sum_{a}f\left(
a\right) \,P_{a}SP_{a}.
\label{eq:Qhat}
\end{eqnarray}
This will be our quantum mechanical analogue of the Baez-Fong map (\ref{eq:BFhat}): where
they convert random variables on a $d$-dimensional sample space into a
diagonal matrix, we convert a hermitean matrix into a CP
map. We, in fact, see that
\begin{eqnarray*}
\mathrm{tr}\left\{ \widehat{f\left( A\right) }\left( \varrho \right) \right\} 
&=&\sum_{a}f\left( a\right) \,\mathrm{tr}\left\{ P_{a}\varrho P_{a}\right\}  \\
&=&\sum_{a}f\left( a\right) \,\mathrm{tr}\left\{ \varrho P_{a}\right\} =\mathrm{tr}\left\{
f\left( A\right) \,\varrho \right\} \\
&=& \mathbb{E}\left[ f\left( A\right) %
\right] .
\end{eqnarray*}

Now let us consider the criterion
\begin{eqnarray}
\left[ \widehat{f\left( A\right) },\mathcal{T}_{t}\right] =0,\quad \text{for all }%
t\geq 0,  \label{eq:QuantumBF}
\end{eqnarray}
where the commutator of maps is now understood as $\left[ \mathcal{T}_{1},\mathcal{T}_{2}\right]
\equiv \mathcal{T}_{1}\circ \mathcal{T}_{2}-\mathcal{T}_{2}\circ \mathcal{T}_{1}$.

\bigskip

\noindent \textbf{Proposition}
\textit{The condition (\ref{eq:QuantumBF}) implies that $A$ is a constant for
the quantum Markov dynamics.}

\textit{proof:}
If the condition (\ref{eq:QuantumBF}) is satisfied, then we have that
\begin{eqnarray*}
\mathbb{E}_{t}\left[ f\left( A\right) \right]  &=&\mathrm{tr}\left\{ f\left( A\right)
\,\mathcal{T}_{t}(\varrho )\right\}  \\
&=&\mathrm{tr}\left\{ \widehat{f\left( A\right) }\circ \mathcal{T}_{t}\left( \varrho \right)
\right\}  \\
&=&\mathrm{tr}\left\{ \mathcal{T}_{t}\circ \widehat{f\left( A\right) }\left( \varrho \right)
\right\}  \\
&=&\mathrm{tr}\left\{ \widehat{f\left( A\right) }\left( \varrho \right) \right\} 
\text{, (as }\mathcal{T}_{t}\text{ is trace-preserving!)} \\
&=&\mathbb{E}\left[ f\left( A\right) \right] . \quad \square
\end{eqnarray*}

For the class of quantum Markov dynamics considered here, it is well-known that they are generated by
infinitesimal maps $\mathcal{M} : \mathfrak{A} \mapsto \mathfrak{A}$ so that
\begin{eqnarray*}
\mathcal{T}_t = e^{t \mathcal{M}},
\end{eqnarray*}
where we have the specific form
\begin{eqnarray}
\mathcal{M}( S) &\equiv& \sum_k \left\{ L_k S L_k^\ast -\frac{1}{2} SL^\ast_k L_k - \frac{1}{2} L^\ast_k L_k S \right\} + i [S,H],
\label{eq:quantum_M}
\end{eqnarray}
with $\{L_k \}$ a collection of operators in $\mathfrak{A}$, and $H\in \mathfrak{A}$ Hermitean. The state $\varrho_t = \mathcal{T}_t (\varrho )$ then satisfies the master equation
$ \frac{d}{dt} \varrho_t = \mathcal{M}(\varrho_t )$. 

The infinitesimal form of the condition (\ref{eq:QuantumBF}) is then
\begin{eqnarray}
\left[ \widehat{f\left( A\right) },\mathcal{M}\right] =0 .
\end{eqnarray}

We remark that $\mathcal{T}_t$ is the Schr\"{o}dinger picture form of the channel - however, we can also work with the Heisenberg picture form $\mathcal{J}_t$. These are CP maps
determined by the duality $ \mathrm{tr} \{ \mathcal{T}_t (S) \, A \} \equiv \mathrm{tr} \{S \, \mathcal{J}_t (A) \}$, 
for all $S,A \in \mathfrak{A}$. The trace-preserving property of $\mathcal{T}_t$ is equivalent to the property $\mathcal{J}_t (I_d ) = I_d$. The generator of the semigroup $\mathcal{J}_t$ is then the Gorini-Kossokowski-Sudarshan-Lindblad generator $\mathcal{L}$ adjoint to $\mathcal{M}$ which takes the form\cite{Alicki_Fannes}
\begin{eqnarray}
\mathcal{L}( A) &\equiv& \sum_k \left\{ L_k^\ast A L_k -\frac{1}{2} AL^\ast_k L_k - \frac{1}{2} L^\ast_k L_k A \right\} - i [A,H],
\label{eq:quantum_L}
\end{eqnarray}

An observable $A$ is said to be a fixed point of a quantum Markov dynamics if it is a fixed point of the corresponding Heisenberg maps, that is
$\mathcal{J}_t (A) =A$ for all $t \geq 0$. Clearly, if $A$ is a fixed point of $\mathcal{J}_t$ then $ \mathrm{tr} \{ \mathcal{T}_t (\varrho) \, A \} 
\equiv \mathrm{tr} \{\varrho \, \mathcal{J}_t (A) \} = \mathrm{tr} \{ \varrho \, A \} $, so it is a constant of the quantum Markov dynamics.

\noindent \textbf{Proposition}
\textit{An observable $A$ satisfies condition (\ref{eq:QuantumBF}) if and only if it is a fixed point of the corresponding Heisenberg maps.}

 \textit{proof:}
 We begin by noting that
\begin{eqnarray*}
\mathrm{tr} \{ [ \widehat{f(A)}, \mathcal{T}_t ] (S) \} &=& \mathrm{tr} \{ f(A) \mathcal{T}_t (S) \} - \mathrm{tr} \{ \widehat{f(A) } (S) \} \\
&=& \mathrm{tr} \{ S \,  \mathcal{J}_t (A) \} - \mathrm{tr} \{  S \, f(A)   \} ,
\end{eqnarray*}
and so if $A$ is a fixed point of $\mathcal{J}_t$, then $[ \widehat{f(A)}, \mathcal{T}_t ]=0$ as $S$ was arbitrary.

Conversely suppose that condition (\ref{eq:QuantumBF}) is satisfied. Let us take the Kraus form for map at time $t$:
$\mathcal{T}_t (S) \equiv \sum_k V_k(t) S V_k^\ast$. (Note that $V_k (t)$ is not required to depend continuously on $t$.) 
The Heisenberg map will then have the adjoint form $\mathcal{J}_t (A) =\sum_k V_k(t)^\ast A V_k(t) $.
Then the condition implies that
\begin{eqnarray*}
\sum_{k,a} V_k(t) P_a SP_a V_k (t)^\ast \, f(a) =\sum_{k,a} f(a) \,  P_a V_k(t) S V_k(t)^\ast P_a.
\end{eqnarray*}
As the choice of $f$ was arbitrary, this implies that 
\begin{eqnarray}
\sum_{k} V_k(t) P_a SP_a V_k (t)^\ast  =\sum_{k} P_a V_k(t) S V_k(t)^\ast P_a.
\label{eq:test}
\end{eqnarray}
If we set $ S=P_b$ where $a \neq b$, then we find that (\ref{eq:test}) reduces to $0=\sum_k (P_a V_k (t) P_b) (P_b V_k(t)^\ast P_a)$ and so we see that
$P_a V_k(t) P_b \equiv 0$. On the other hand, setting $S=P_a$ leads to the conclusion that if $Y_a (t)= \sum_k V_k (t) P_a V_k(t)^\ast$, then $Y_a (t) = P_a Y_a (t) P_a$.
It follows that
\[
[V_k(t) , P_a] =0
\]
for each eigenvalue $a$ of $A$, and therefore $[V_k (t) ,A] =0$. As $A$ commutes with all the Kraus operators, we get that
\[
\mathcal{J}_t (A) =\sum_k V_k(t)^\ast A V_k(t) \equiv \sum_k V_k(t)^\ast V_k(t)\, A  =A.
\]
So condition (\ref{eq:QuantumBF}) implies that $A$ is a fixed point of the quantum Markov dynamics.
$\square$

\section{Constants of Quantum Markov Semigroups (General Case)}
In this section, we investigate the question of characterising the constant observables for a given quantum dynamic semigroup in the general setting where the
observables belong to the set $\mathfrak{A}$ of bounded operators on a fixed separable Hilbert space $\mathfrak{h}$. Let $\mathfrak{T} ( \mathfrak{h} )$ denote
the set of traceclass operators on $\mathfrak{h}$ then the constants of a QDS $(\mathcal{T}_t )_{t\geq 0}$ are the elements of the set
\[
\mathfrak{M} = \bigg\{ A \in \mathfrak{A} : \mathrm{tr} \{ (\mathcal{T}_t \varrho ) \, f( A) \} = \mathrm{tr} \{ \varrho \, A \} , 
\; \forall \varrho \in \mathfrak{T} ( \mathfrak{h}  ),\
t\geq 0 , f \in \mathscr{C} (\mathbb{C} ) \bigg\}.
\]
Note that $\mathfrak{M}$ may contain no-self-dajoint elements, and may be a non-commuting set of operators.

Transferring to the Heisenberg picture, we likewise define the \textit{fixed points} of the QDS 
$( \mathcal{J}_t )_{t\geq 0}$ to be the collection of operators
\begin{eqnarray*}
\mathfrak{F} \triangleq \bigg\{ A \in \mathfrak{A} : \mathcal{J}_t (A) = A, \, \forall \,  t \geq 0 \bigg\}.
\end{eqnarray*}

Intuitively, one suspects for any reasonably defined QDS the constants of the QDS $\mathfrak{M}$ and the fixed points of the QDS $\mathfrak{F}$ are the 
same. To make any progress in the general case however we need to meet certain technical requirements. The following assumption will turn out to be sufficient.

\bigskip

\noindent \textbf{Postulate (P)}
\textit{There exists a strictly positive density matrix $\hat \varrho >0$ on $\mathfrak{h}$ that is stationary with respect to QDS, that is $\mathcal{T}_t (\hat \varrho )=
\hat \varrho$ for all $t>0$ where $\mathcal{T}_t$ is the Schr\"{o}dinger picture form of the QDS.}

\bigskip

We can now state a Noether Theorem for Quantum Dynamical Semigroups.

\bigskip

\noindent \textbf{Theorem }
\textit{
Suppose we are given a QDS of maps $(\mathcal{T}_t )_{t \ge 0}$ continuous in the trace-norm topology. If the postulate \textbf{(P)} holds then the collection of fixed points $\mathfrak{F}$ forms a von Neumann algebra and this is the
algebra $\mathfrak{M}$ of constants for the QDS. Moreover, the infinitesimal generator then takes the form (\ref{eq:quantum_M}) with $H, L_k ,\sum_k L^\ast_k L_k \in \mathfrak{A}$
and we have}
\begin{eqnarray}
\mathfrak{M} \equiv \bigg\{ A \in \mathfrak{A} : [A,H]=0,\,  [A,L_k ]=0, [A,L_k^\ast ]=0 , \forall \,  k\bigg\} . 
\label{eq:VF}
\end{eqnarray}

\textit{proof:}
We first observe that the dual Heisenberg semigroup $(\mathcal{J}_t )_{t \geq 0}$ will be norm continuous.
For $A$ to be a constant under the QDS we require that $ \mathrm{tr} \{ \varrho \, \mathcal{J}_t ( f(A) ) \}=\mathrm{tr} \{ \varrho \, f(A)  \}$ for each $t>0$ and each density matrix $\varrho$, and any bounded continuous function $f$, say. Consequently, we must have that $f(A)$ will also be a fixed point of the QDS. What is not immediate at present is
the property that $A\in \mathfrak{F}$ necessarily implies that $f(A) \in \mathfrak{F}$, or in other words that the fixed points form an algebra.

However, the existence of $\hat \varrho$ from the postulate implies that the state $\hat{ \mathbb{E} }[ \, \cdot \, ]$ it generates is a normal, faithful stationary state for the QDS \cite{faithful}.
This is ensures that the fixed points $\mathfrak{F}$ now form a von Neumann algebra \cite{Fagnola_Rebolledo}. Consequently, if $A$ is a fixed point, then any continuous bounded function $f(A)$ is also a fixed point, and therefore $A$ is a constant of the QDS.

The generators of norm-continuous QDS have Lindblad generators of the form (\ref{eq:quantum_M}).
The characterisation of the fixed points as given in (\ref{eq:VF}) then follows from Theorem 3.3, p.281, of Frigerio and Verri \cite{FV}.
$\square$

\bigskip

Note that $\mathfrak{M}$ may be non-commutative. The requirement that the Schr\"{o}dinger semigroup $(\mathcal{T}_t )_{t \geq 0}$ be continuous in the
trace-norm topology implies that the infinitesimal generator $\mathcal{M}$ still takes the form (\ref{eq:quantum_M}). This is the classic result
of Lindblad \cite{Lindblad}. This restriction is not essential, and sufficient conditions for when the form (\ref{eq:quantum_M}) holds but with the operators $H, L_k$ are unbounded
have been derived by Fagnola and Rebolledo \cite{Fagnola_Rebolledo}, see also \cite{Fagnola_Rebolledo_98}.

Postulate \textbf{(P)} however is more essential as it ensures that the fixed points $\mathfrak{M}$
form a von Neumann algebra. We may explain the characterisation (\ref{eq:VF}) of $\mathfrak{M}$ in terms of the infinitesimal generator as follows: for a closed Hamiltonian evolution, the commutativity requirement with the Hamiltonian $H$ is straightforward; for $A$ self-adjoint to be a constant, we would require $\mathcal{J}_t (A^2) \equiv \mathcal{J}_t (A)^2$ which at the infinitesimal level implies $\mathcal{L} (A^2) - \mathcal{L} (A)\, A-A \, \mathcal{L} (A) =0$, but this last expression takes the form $\sum_k [A,L_k ]^\ast [A,L_k ] \equiv 0$ and this requires $[A,L_k ]=0$ for each $k$. Every non-self-adjoint Noether constant should then be the sum $A_1 +i A_2$ with $A_1,A_2$ self-adjoint Noether constants.

Postulate \textbf{(P)} deserves some further comment as it implies that $\mathfrak{M}$ is invariant
under the modular group $(\sigma^{\hat{ \mathbb{E}} }_t )_{t \in \mathbb{R}}$ from  the Tomita-Talesaki theory, in fact,
\[
[ \sigma^{\hat{ \mathbb{E}} }_t  , \mathcal{J}_t ] =0 ,
\]
and moreover that there exists a conditional expectation $ \hat{\mathbb{E} }[\cdot | \mathfrak{M}] $
from $\mathfrak{A}$ onto $\mathfrak{M}$. In the present situation, $\hat{\mathbb{E}} [\cdot | \mathfrak{M}]$ will be
the (unique) faithful, normal conditional expectation with $\hat{\mathbb{E} }[\cdot ]$ as invariant state:
\[
\hat{\mathbb{E}} [M \hat{\mathbb{E}} [A| \mathfrak{M}]] = \hat{\mathbb{E} } [MA],
\]
for all $M \in \mathfrak{M}$.
We remark, that conditional expectations from a von Neumann algebra onto a sub-algebra do not generally exist.

The original motivation for quantum dynamical semigroups was to 
study convergence to equilibrium of arbitrary initials states for general quantum open systems. The results in this area may be rephrased using the Noether Theorem
as saying that all states converge to the (unique) equilibrium state $\hat \varrho$ if and only if the only constants of the QDS are multiples of identity,
that is $\mathfrak{M} \equiv \mathbb{C} \, I$. More generally, a non-trivial algebra $\mathfrak{M}$ of observables that are constant under a QDS.

A general condition for the existence of a stationary state $\hat \varrho >0$ are given by Fagnola and Rebolledo \cite{Fagnola_Rebolledo}, 
including their Theorem 4.2 which gives the condition on the infinitesimal generator. 

We finally remark that $\hat{\mathbb{E}} [\mathcal{J}_t (A) | \mathfrak{M}] = \mathcal{J}_t (\hat{\mathbb{E}} [A| \mathfrak{M}] )$ as both sides
equal $\hat{\mathbb{E}} [A| \mathfrak{M}]$ due to the inavriance and stationarity of the state $\hat{\mathbb{E}} $. This may be written as
\begin{eqnarray}
\bigg[ \hat{\mathbb{E}} [\cdot| \mathfrak{M}] , \mathcal{J}_t \bigg] =0, \quad \forall t\geq 0. 
\end{eqnarray}
This is the quantum analogue of the criterion established by Baez and Fong for classical Markov semigroups.

\begin{acknowledgments}
JG acknowledges support from EPSRC grant EP/L006111/1 Quantum Stochastic Analysis For Nano-photonic Circuit Design. OGS acknowledges Russian Foundation for Basic
Research grant number 14-01-00516. Both JG and OGS are grateful to the hospitality of the Bernoulli Interfacultary Centre, EPF Lausanne, where this research was carried out. TSV is partially supported by NCCR
SwissMAP and grant 200021-140238, both of the Swiss National Science Foundation.
\end{acknowledgments}

%$ \, $

%\bibliography{Noether}

\begin{thebibliography}{9}
\bibitem{Baez_Fong}  J.C. Baez, B. Fong, J. Math. Phys., 54:013301 (2013).

\bibitem{Alicki_Fannes} R. Alicki and M. Fannes, Quantum Dynamical Systems, Oxford University Press, Oxford (2000).


\bibitem{sigma-algebra} Actually since $\Omega$ is finite at present, we only need the algebra of subsets: specifically $\mathscr{M}$ is the collection of sets
consisting of the empty set, each of the equivalence classes, and all of their possible unions. However, the general result requires a $\sigma$-algebra.


\bibitem{Fagnola_Rebolledo} F. Fagnola, R. Rebolledo, 
in Open Quantum Systems III. Recent Developments, Lecture Notes in Mathematics Volume \textbf{1882}, pp 161-205,  (2006)


\bibitem{faithful} The terminology here is standard in operator theory: a state $\hat{\mathbb{E}}$ is normal if it takes the form
$ \mathrm{tr} \{ \varrho \, \cdot \} $, and faithful if the density matrix $\hat \varrho$ is trictly positive (that is $\hat{ \mathbb{E}}
[X ]=0$ for a positive operator $X$ if and only if $X=0$).

\bibitem{FV} A. Frigerio and M. Verri, Math. Zeitschrift \textbf{180}, no. 2, pp. 275–286, (1982)

\bibitem{Lindblad} G. Lindblad,
%On the generators of quantum dynamical semigroups,
Commun. Math. Phys., \textbf{48}, pp. 119-130 (1976)

\bibitem{Fagnola_Rebolledo_98} F. Fagnola, R. Rebolledo, 
%The approach to equilibrium of a class of quantum dynamical semigroups,
Inf. Dim. Anal. Q. Prob. and Rel. Topics, \textbf{1} (4), pp. 1-12 (1998)

\end{thebibliography}

\end{document}